\documentclass{article} 
\usepackage{nips14submit_e,times}
\usepackage{url}

\usepackage{natbib}

\title{Lazy ABC}

\author{
Dennis Prangle \\
Mathematics and Statistics Department\\
University of Reading\\
Reading, UK \\
\texttt{d.b.prangle@reading.ac.uk} \\
}

\nipsfinalcopy 

\usepackage{bbm} 
\usepackage{amssymb} 
\usepackage{amsmath} 
\DeclareMathOperator{\E}{E}

\usepackage{float} 
\newfloat{algorithm}{htbp}{loa} 
\floatname{algorithm}{Algorithm}
\newcommand{\algline}{\rule{\textwidth}{0.2mm}}

\begin{document}

\maketitle

\begin{abstract}
ABC algorithms involve a large number of simulations from the model of interest,
which can be very computationally costly.
This paper summarises the lazy ABC algorithm of \cite{Prangle:2014}, which reduces the computational demand by abandoning many unpromising simulations before completion.
By using a random stopping decision and reweighting the output sample appropriately, the target distribution is the same as for standard ABC.
Lazy ABC is also extended here to the case of non-uniform ABC kernels,
which is shown to simplify the process of tuning the algorithm effectively.
\end{abstract}

\section{Algorithms}

Approximate Bayesian computation (ABC) approximates Bayesian inference on parameters $\theta$ with prior $\pi(\theta)$ given data $y_{\text{obs}}$.
It must be possible to simulate data $y$ from the model of interest given $\theta$.
This implicitly defines a likelihood function $L(\theta)$: the density of $y_{\text{obs}}$ conditional on $\theta$.

A standard importance sampling version of ABC samples parameters $\theta_{1:N}$ from an importance density $g(\theta)$ and simulates corresponding datasets $y_{1:N}$.
Weights $w_{1:N}$ are calculated by equation \eqref{eq:ABCweight} below.
Then for a generic function $f(\theta)$, an estimate of its posterior expectation $\E[f(\theta) | y_{\text{obs}}]$ is $\mu_f = [\sum_{i=1}^N w_i]^{-1} \sum_{i=1}^N f(\theta_i) w_i$.
An estimate of the normalising constant $\pi(y_{\text{obs}}) = \int \pi(\theta) L(\theta) d\theta$, used in Bayesian model choice, is $\hat{z} = N^{-1} \sum_{i=1}^N w_i$.
Under the ideal choice of weights, $w_i = L(\theta_i) \pi(\theta_i) / g(\theta_i)$, these estimates converge (almost surely) to the correct quantities as $N \to \infty$ \cite{Liu:1996}.
In applications where $L(\theta)$ cannot be evaluated ABC makes inference possible with the trade-off that it gives approximate results.
That is, the estimators converge to approximations of the desired values.

The ABC importance sampling weights avoid evaluating $L(\theta)$ by using:
\begin{align} 
w_{\text{ABC}} &= L_{\text{ABC}} \pi(\theta) / g(\theta) \label{eq:ABCweight} \\
\text{where} \qquad L_{\text{ABC}} &= K[d(s(y), s(y_{\text{obs}})) / h] \label{eq:LABC}
\end{align}
Here:
\begin{itemize}
\item
$L_{\text{ABC}}$ acts as an estimate (up to proportionality) of $L(\theta)$.
This and $w_{\text{ABC}}$ are random variables since they depend on $y$, a random draw from the model conditional on $\theta$.
\item
$s(\cdot)$ maps a dataset to a lower dimensional vector of \emph{summary statistics}.
\item
$d(\cdot,\cdot)$ maps two summary statistic vectors to a non-negative value. This defines the \emph{distance} between two vectors.
\item
$K[\cdot]$, the \emph{ABC kernel} maps from a non-negative value to another. A typical choice is a uniform kernel $K[x]=\mathbbm{1}(x \in [0,1])$, which makes an accept/reject decision. Another choice is a normal kernel $K[x]=e^{-x^2}$.
\item
$h \geq 0$ is a tuning parameter, the \emph{bandwidth}. It controls how close a match of $s(y)$ and $s(y_{\text{obs}})$ is required to produce a significant weight.
\end{itemize}
The interplay between these tuning choices has been the subject of considerable research but is not considered further here.
For further information on this and all aspects of ABC see the review papers \cite{Beaumont:2010, Marin:2012}.

Lazy ABC splits simulation of data into two stages.
First the output of some \emph{initial simulation stage} $x$ is simulated conditional on $\theta$,
then, sometimes, a full dataset $y$ is simulated conditional on $\theta$ and $x$.
The latter is referred to as the \emph{continuation simulation stage}.
The variable $x$ should encapsulate all the information which is required to resume the simulation so may be high dimensional.
There is considerable freedom of what the initial simulation stage is.
It may conclude after a prespecified set of operations, or after some random event is observed.
Another tuning choice is introduced, the \emph{continuation probability} function $\alpha(\theta, x)$.
This outputs a value in $[0,1]$ which is the probability of continuing to the continuation simulation stage.
The desired behaviour in choosing the initial simulation stage and $\alpha$ is that simulating $x$ is computationally cheap but can be used to save time by assigning small continuation probabilities to many unpromising simulations.

Given all the above notation, lazy ABC is Algorithm \ref{alg:lazyABC}.
To avoid division by zero in step 5, it will be required that $\alpha(\theta,x)>0$, although this condition can be weakened \cite{Prangle:2014}.

\begin{algorithm}[htp]
\algline \\
{\bf Algorithm:}\\
Perform the following steps for $i=1,\ldots,N$:
\begin{enumerate}
\item[1] Simulate $\theta_i$ from $g(\theta)$.
\item[2] Simulate $x_i$ conditional on $\theta_i$ and set $a_i = \alpha(\theta_i, x_i)$.
\item[3] With probability $a_i$ continue to step 4. Otherwise perform \emph{early stopping}: let $\ell_i=0$ and go to step 6.
\item[4] Simulate $y_i$ conditional on $\theta_i$ and $x_i$.
\item[5] Set $\ell_i = K[d(s(y_i), s(y_{\text{obs}})/h)] / a_i$.
\item[6] Set $w_i = \ell_i \pi(\theta_i) / g(\theta_i)$.
\end{enumerate}
{\bf Output:}\\ A set of $N$ pairs of $(\theta_i, w_i)$ values.\\
\algline
\caption{Lazy ABC \label{alg:lazyABC}}
\end{algorithm}

Lazy ABC has the same target as standard ABC importance sampling,
in the sense that the Monte Carlo estimates $\mu_f$ and $\hat{z}$
converge to the same values for $N \to \infty$.
This is proved by Theorem 1 and related discussion in \cite{Prangle:2014}.
A sketch of the argument is as follows.
Standard ABC is essentially an importance sampling algorithm:
each iteration samples a parameter value $\theta$ from $g(\theta)$ and assigns it a random weight $w$ given by \eqref{eq:ABCweight}.
The randomness is due to the random simulation of data $y$.
The expectation of this weight conditional on $\theta$ is
\[
\E [w_{\text{ABC}} | \theta ] = \E [ L_{\text{ABC}} | \theta ] \pi(\theta)/g(\theta)
\]
where expectation is taken over values of $y$.

Lazy ABC acts similarly but uses different random weights
\begin{equation} \label{eq:wlazyABC}
w_{\text{lazy}} = \begin{cases}
L_{\text{ABC}} a^{-1} \pi(\theta)/g(\theta) & \text{with probability } a=\alpha(\theta,x) \\
0 & \text{otherwise}
\end{cases}
\end{equation}
The randomness here is due to simulation of $x$ and $y$.
Taking expectations gives:
\begin{align*}
\E[w_{\text{lazy}} | \theta, x] &= \E[L_{\text{ABC}} | \theta, x] \pi(\theta)/g(\theta) \\
\Rightarrow \qquad \E[w_{\text{lazy}} | \theta] &= \E[L_{\text{ABC}} | \theta] \pi(\theta)/g(\theta)
\end{align*}
From the theory of importance sampling algorithms with random weights (see \cite{Prangle:2014}) this ensures that both algorithms target the same distribution.

This argument shows lazy ABC targets the same $\mu_f$ and $\hat{z}$ quantities as standard ABC,
for any choice of initial simulation stage and $\alpha$.
However, for poor choices of these tuning decisions it may converge very slowly.
The next section considers effective tuning.

\section{Lazy ABC tuning} \label{sec:tuning}

The quality of lazy ABC tuning can be judged by an appropriate measure of \emph{efficiency}.
Here this is defined as effective sample size (ESS) divided by computing time.
The ESS for a sample with weights $w_1, \ldots, w_N$ is
\[
N_{\text{eff}} = N \bigg[ N^{-1}\sum_{i=1}^N w_i \bigg]^2 / \bigg[ N^{-1} \sum_{i=1}^N w_i^2 \bigg].
\]
It can be shown \cite{Liu:1996} that for large $N$ the variance of $\mu_f$ typically equals that of $N_{\text{eff}}$ independent samples.
Computing time is taken to be the sum of CPU time for each core used (as the lazy ABC iterations can easily be performed in parallel.)

Theorem 2 of \cite{Prangle:2014} gives the following results on the choice of $\alpha$ which maximises the efficiency of lazy ABC in the asymptotic case of large $N$.
For now let $\phi$ represent $(\theta, x)$.
Then the optimal choice of $\alpha$ is of the following form:
\begin{align}
\alpha(\phi) &= \min\left\{1, \lambda \left[\frac{ \gamma(\phi) }{\bar{T}_2(\phi)}\right]^{1/2} \right\} \label{eq:tuning} \\
\text{where} \qquad \gamma(\phi) &= \E \big[ L_{\text{ABC}}^2 \pi(\theta)^2 g(\theta)^{-2} | \phi \big]. \label{eq:gamma}
\end{align}
Here $\gamma(\phi)$ is the expectation given $\phi$ of $w_{\text{ABC}}^2$, the squared weight which would be achieved under standard ABC importance sampling;
$\bar{T}_2(\phi)$ is the expected time for steps 4-6 of Algorithm \ref{alg:lazyABC} given $\phi$;
$\lambda \geq 0$ is a tuning parameter that controls the relative importance of maximising ESS (maximised by $\lambda=\infty$) and minimising computation time (minimised by $\lambda=0$).

A natural approach to tuning $\alpha$ in practice is as follows. The remainder of the section discusses these steps in more detail.
\begin{enumerate}
\item Using Algorithm \ref{alg:lazyABC} with $\alpha \equiv 1$ simulate training data $(\theta_i, x_i, y_i, t^{(1)}_i, t^{(2)}_i)_{1 \leq i \leq M}$. Here $t^{(1)}_i$ is the time to perform steps 1-3 of Algorithm \ref{alg:lazyABC} and $t^{(2)}_i$ is the time for steps 4-6.
\item Estimate $\gamma(\phi)$ and $\bar{T}_2(\phi)$ from training data.
\item Choose $\lambda$ to maximise an efficiency estimate based on the training data.
\item Decide amongst various choices of initial simulation stage (and $\phi$, see below) by maximising estimated efficiency. By collecting appropriate data for these choices in step 1 it is not necessary to repeat it.
\end{enumerate}
Step 2 is a regression problem, but is not feasible for $\phi=(\theta,x)$ as this will typically be very high dimensional.
Instead $\alpha$ can be based on low dimensional features of $(\theta,x)$, referred to as \emph{decision statistics}.
That is, only $\alpha$ functions of the form $\alpha(\phi(\theta,x))$ are considered, where $\phi(\theta,x)$ outputs a vector of decision statistics.
The optimal such $\alpha$ is again given by \eqref{eq:tuning} and \eqref{eq:gamma}.
The choice of which decision statistics to use can be included in step 4 above.

Estimating $\gamma(\phi)$ by regression is also challenging if there are regions of $\phi$ space for which most of the responses are zero.
This is typically the case for uniform $K$.
In \cite{Prangle:2014} various tuning methods were proposed for uniform $K$ but these are complicated and rely on strong assumptions.
A simpler alternative used here is to use a normal $K$ as it has full support.

Local regression techniques \cite{Hastie:2009} are suggested for step 2.
This is because the behaviour of the responses typically varies considerably for different $\phi$ values, motivating fitting separate regressions.
Firstly, the typical magnitude of $L_{\text{ABC}}$ varies over widely different scales.
Secondly, for both regressions the distribution of the residuals may also vary with $\phi$. 
To ensure positive predictions, the use of degree zero regression is suggested i.e.~a Nadaraya-Watson kernel estimator.

The efficiency estimate required in steps 3 and 4 can be formed from the training data and proposed choice of $\alpha$.
Let $(\alpha_i)_{1 \leq i \leq M}$ be the $\alpha$ values for the training data
and $(l_i)_{1 \leq i \leq M}$ be the values of $L_{\text{ABC}}$.
The realised efficiency of the training data is not used since it is based on a small sample size.
Instead the asymptotic efficiency is estimated.
Under weak assumptions (see \cite{Prangle:2014}) this is $\E(w)^2 \E(w^2)^{-1} \E(T)^{-1}$,
where random variable $T$ is the CPU time for a single iteration of lazy ABC.
Note that $E(w)$ is constant (the ABC approximation for the normalising constant $\pi(y_{\text{obs}})$) under any tuning choices, so it is omitted.
This leaves an estimate up to proportionality of $[\E(w^2) \E(T)]^{-1}$ which can be used to calculate efficiency relative to standard ABC (found by setting $\alpha \equiv 1$).
An estimate of $\E(T)$ is
$\hat{T} = M^{-1} [\sum_{i=1}^{M} t^{(1)}_i + \sum_{i=1}^{M} \alpha_i t^{(2)}_i]$.
Using \eqref{eq:wlazyABC} an estimate of $E(w^2)$ is
$\widehat{w^2} = M^{-1} \sum_{i=1}^{M} l_i^2 \alpha_i^{-1} \pi(\theta_i)^2 g(\theta_i)^{-2}$

\section{Example}

As an example the spatial extremes application of \cite{Erhardt:2012} is used.
This application and the implementation of lazy ABC is described in full in \cite{Prangle:2014}.
A short sketch is that the model of interest has two parameters $(c,\nu)$.
Given these, data $y_{t,d}$ can be generated for years $1 \leq t \leq 100$ and locations $1 \leq d \leq 20$.
These represent annual extreme measurements e.g.~of rainfall or temperature.
An ABC approach has been proposed including choices of $s(\cdot)$ and $d(\cdot,\cdot)$.
Also, given data for a subset of locations an estimate of the ABC distance can be formed.

Simulation of data is hard to interrupt and later resume.
However the most expensive part of the process is calculating the summary statistics,
which involves calculating certain coefficients for every triple of locations.
Therefore the initial simulation stage of lazy ABC is to simulate all the data and calculate an estimated distance based on a subset of locations $L$, which is used as the decision statistic $\phi$.
The continuation stage is to calculate the coefficients for the remaining triples and return the realised distance.

Tuning of lazy ABC was performed as described in Section \ref{sec:tuning}, using backwards selection in step 4 to find an appropriate subset of locations to use as $L$.
To fit the regressions estimating $\gamma(\phi)$ and $\bar{T}_2(\phi)$ a Nadaraya-Watson kernel estimator was used with a Gaussian kernel and bandwidth 0.5, chosen manually.

Repeating the example of \cite{Prangle:2014}, 6 simulated data sets were analysed using standard and lazy ABC.
Each analysis used $10^6$ simulations in total.
In lazy ABC $M=10^4$ of these were used for training.
The results are shown in Table \ref{tab:SErep}.
The efficiency improvements of lazy ABC relative to standard ABC are of similar magnitudes to those in \cite{Prangle:2014} but are less close to the values estimated in step 3 of tuning.

\begin{table}[pht]
\centering
\fbox{ \footnotesize
\begin{tabular}{cc|c|cc|cc}
\multicolumn{2}{c|}{Parameters}                   & Standard       & \multicolumn{2}{|c|}{Lazy} & \multicolumn{2}{|c}{Relative efficiency} \\
$c$    & $\nu$                                    & Time ($10^3$s) & Time ($10^3$s) & ESS       & Estimated & Actual \\ 
  \hline                                     
 0.5  & 1                                         & 26.7           & 8.0           & 131.6      & 3.9      & 2.2 \\ 
 1    & 1                                         & 25.6           & 7.1           & 174.2      & 4.5      & 3.1 \\ 
 1    & 3                                         & 25.5           & 8.3           & 185.3      & 3.8      & 2.8 \\ 
 3    & 1                                         & 25.6           & 7.6           & 267.2      & 4.2      & 4.5 \\ 
 3    & 3                                         & 25.2           & 8.2           & 193.5      & 3.9      & 3.0 \\ 
 5    & 3                                         & 25.7           & 8.4           & 162.4      & 3.7      & 2.5
\end{tabular}}
\caption{\label{tab:SErep}Simulation study on spatial extremes.  Each row represents the analysis of a simulated dataset under the given values of parameters $c$ and $\nu$. In each analysis a choice of $\epsilon$ was made under standard ABC so that the ESS was 200, and the same value was used for lazy ABC. The lazy ABC output sample includes the training data, as described in \cite{Prangle:2014}. Also its computation time includes the time for tuning calculation (roughly 70 seconds). Iterations were run in parallel and computation times are summed over all cores used.}
\end{table}

\section{Conclusion}

The paper has reviewed lazy ABC \cite{Prangle:2014}, a method to speed up ABC without introducing further approximations to the target distribution.
Unlike \cite{Prangle:2014}, non-uniform ABC kernels have been considered.
This allows a simpler approach to tuning, which provides a comparable three-fold efficiency increase in a spatial extremes example.

Several extensions to lazy ABC are described in \cite{Prangle:2014}:
multiple stopping decisions, choosing $h$ after running the algorithm and a similar scheme for likelihood-based inference.
Other potential extensions include using the lazy ABC approach in ABC versions of MCMC or SMC algorithms, or focusing on model choice.

\newpage

\bibliographystyle{plainnat}
\bibliography{lazy}

\end{document}